# DESIGN AND IMPLEMENTATION OF REAL TIME AES-128 ON REAL TIME OPERATING SYSTEM FOR MULTIPLE FPGA COMMUNICATION

Rourab Paul[1]
Dept. of Electronic Science[1]
Kolkata, India
rourabpaul@gmail.com
University of Calcutta

Sangeet Saha[2]
Dept. of Computer Science and Engineering[2]
Kolkata, India
sangeet.saha87@gmail.com
University of Calcutta

Suman Sau[3]
A. K. Choudhury School of Information Technology[3]
Kolkata, India
sumansau@gmail.com
University of Calcutta

Amlan Chakrabarti[4]
A. K. Choudhury School of Information Technology[4]
Kolkata, India
acakcs@caluniv.ac.in
University of Calcutta

*Abstract-* **Security is the most important part in data communication system, where more randomization in secret keys increases the security as well as complexity of the cryptography algorithms. As a result in recent dates these algorithms are compensating with enormous memory spaces and large execution time on hardware platform. Field programmable gate arrays (FPGAs), provide one of the major alternative in hardware platform scenario due to its reconfiguration nature, low price and marketing speed. In FPGA based embedded system we can use embedded processor to execute particular algorithm with the inclusion of a real time operating System (RTOS), where threads may reduce resource utilization and time consumption. A process in the runtime is separated in different smaller tasks which are executed by the scheduler to meet the real time dead line using RTOS. In this paper we demonstrate the design and implementation of a 128-bit Advanced Encryption Standard (AES) both symmetric key encryption and decryption algorithm by developing suitable hardware and software design on Xilinx Spartan- 3E (XC3S500E-FG320) device using an Xilkernel RTOS, the implementation has been tested successfully The system is optimized in terms of execution speed and hardware utilization.**

## I. INTRODUCTION

Data security is an essential objective for the military and diplomatic services which have many commercial uses and applications such as electronic banking, electronic mail, internet network service, messaging networks etc. As an efficient and cost-effective cryptographic algorithm AES [1] algorithm has broad applications, including smart cards and cellular phones, WWW servers and automated teller machines (ATMs). Establishing reliable communication between multiple FPGA systems/cards is an essential component for developing complex real time systems used for applications like real time data acquisition and processing [2, 3]. Though related work exists for FPGA based AES implementation [4,5,6,7,8,9,10,11,12].But there exist no implementation based on real time data encryption and decryption over a communication interface between multiple FPGA systems using an RTOS, where each thread is capable of running the encryption and decryption individually.

Our proposed work is an FPGA based design and implementation of the AES-128 algorithm on Real Time Operating System. We have successfully established a secured link between two FPGA systems through RS232 link and have achieved good throughput with a minimum number of resource utilization, compared to the other existing works [13, 14, 15, 16, 17]. The total system functions like a complete system where data are taken from key board in real time by the FPGA, the encryption is done using a thread running on the RTOS. Through the RS232 communication link the cipher data is being transmitted to another FPGA, where the decryption is done by another thread of the RTOS. The decrypted data is visualized on the HyperTerminal for verification. The algorithm and the real time scenario of the system has been analyzed and its hardware utilization proves to be better compared to the related works [13,14,15]. The development platform of our work is Xilinx EDK 11.1 and has an RTOS we have chosen Xilinx owned Xilkernel [18].

The organization of the paper is as follows, Section II describes the design-flow of the AES algorithm for both encryption and decryption process. Section III details out the Real time Operating System on Hardware, Section IV is discussed on proposed hardware architectural design. Section V briefs on the implementation, results and comparison with existing works and the concluding remarks are presented in Section VI.

## II DESIGN OVERVIEW OF AES

The algorithm is composed of three main parts: Cipher, Inverse Cipher and Key Expansion. Cipher converts data, commonly known as plaintext, to an unintelligible form called cipher. Key Expansion generates a key schedule that is used in the Cipher and the Inverse Cipher procedure. Cipher and Inverse Cipher are composed of specific number of rounds (Table 1). For the AES algorithm, the number of rounds to be performed during the execution of the algorithm is dependent on the key length [1].

AES operates on a *4x4* array of bytes (referred to as "state"). The algorithm consists of four different simple operations. These operations are:
- Sub Bytes
- Shift Rows
- Mix Columns
- Add Round Key

TABLE I: FEATURES OF AES FOR DIFFERENT KEY LENGTHS

|  | Block size $N_b$ words | Key length $N_k$ words | Number of rounds $N_r$ |
|---|---|---|---|
| AES– 128_bits key | 4 | 4 | 10 |
| AES-192_bits key | 4 | 6 | 12 |
| AES-256_bits key | 4 | 8 | 14 |

### A. Encryption Process:

The Encryption and decryption process consists of a number of different transformations applied consecutively over the data block bits, in a fixed number of iterations, called rounds. The number of rounds depends on the length of the key used for the encryption process. For key length of 128 bits, the number of iteration required are 10 *($N_r$ = 10)*. As shown in Fig. 1(a), each of the first $N_{r-1}$ rounds consists of 4 Transformations: *SubBytes()*, *ShiftRows()*, *MixColumns()* and *AddRoundKey()*.

*1) Sub Bytes Transformation:* It is a non-linear substitution of bytes that operates independently on each byte of the state using a substitution table (*S* box). This invertible *S*-box is constructed by first taking the multiplicative inverse in the finite field GF (28) with irreducible polynomial $m(x) = x^8 + x^4 + x^3 + x + 1$. The element {00} is mapped to itself. Then affine transformation is applied (over GF (2)).

*2) Shift Rows Transformation:* Cyclically shifts the rows of the state over different offsets. The operation is almost the same in the decryption process except for the fact that the shifting offsets have different values.

*3) Mix Columns Transformation:* This transformation operates on the state column-by-column, treating each column as a four-term polynomial. The columns are considered as polynomials over GF (28) and multiplied by modulo $x^4 + 1$ with a fixed polynomial $a(x) = \{03\} x^3 + \{01\} x^2 + \{01\} x + \{02\}$.

*4) Add Round Key Transformation:* In this transformation, a Round Key is added to the state by a simple bitwise XOR operation. Each Round Key consists of $N_b$ words from the key expansion. Those $N_b$ words are each added into the columns of the state. Key Addition is the same for the decryption process.

*5) Key Expansion:* Each round key is a 4-word (128-bit) array generated as a product of the previous round key, a constant that changes each round, and a series of *S*-Box lookups for each 32-bit word of the key. The Key schedule Expansion generates a total of $N_b *(N_r + 1)$ words.

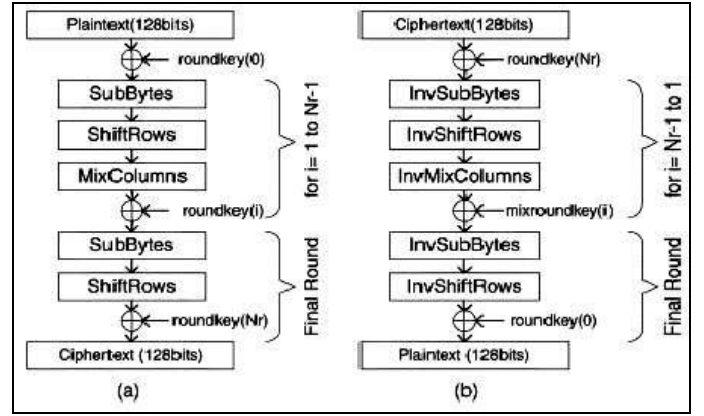

Fig. 1: AES algorithm (a)Encryption structure(b) Equivalent Decryption Structure

### B. Decryption Process:

For decryption, the same process occurs simply in reverse order, taking the 128-bit block of cipher text and converting it to plaintext by the application of the inverse of the four operations. AddRoundKey is the same for both encryption and decryption. However the three other functions have inverses used in the decryption process: Inverse SubBytes, Inverse ShiftRows, and Inverse MixColumns. This process is direct inverse of the encryption process. All the transformations applied in the encryption process are inversely applied to this process. Hence the last round values of both the data and key are first round inputs for the decryption process and follows in decreasing order as shown Fig. 1(b).

This is the brief description of the AES algorithm for details see the reference [1].

## III. REAL TIME OPERATING SYSTEM

REAL – time embedded systems are typically designed for various purposes such as to control or to process data. Characteristics of real-time system include meeting certain deadlines at the right time. To achieve this purpose, real-time operating systems (RTOS) are often used, more specifically an RTOS is a piece of software with a set of APIs for users to develop applications [19] which requires deadline based execution. RTOSes are typically differentiated from generic OSes regarding the following criteria i.e. preemptive or priority-based scheduling, predictability in task synchronization, deterministic behaviors [19]. There are various RTOSes available for microcontroller as well as for FPGA based design i.e. VxWorks, QNX, eCos, LynxOS, and RTLinux [20].

Here we have chosen Xilkernel [18] as our RTOS, which is provided by Xilinx. Xilkernel is a small, robust, and modular kernel. It is highly integrated with the Platform Studio Frame [21]. The advantages for the usage of Xilkerenel as RTOS here, are

- Xilkernel has very low memory footprint, it uses 7-16 kb of BRAM in a multi threaded program [18], which is much smaller than the RTOS used in microcontroller [10].

- In the context of Xilkernel, a *thread* is the unit of execution and is analogous to a process. Threads are coded like functions.

- Xilkernel is structured as a library. The user application source files must link with Xilkernel to access Xilkernel functionality.

- The MicroBlaze kernel requires an external timer to generate periodic interrupts and this is the only hardware requirement that the kernel places.

- If an interrupt controller is present in the system, Xilkernel exports an interrupt handler registering mechanism and invokes the handlers after it pre-processes each hardware interrupt.[18]

Below we shown an outline of the code to implement the Xilkernel on RTOS

```
int main(void)
   {

xilkernel_main();   /* Start the kernel */

/* Control does not reach here */

   }

void* main_thread(void)  /* Statically created first thread */

   {

//Here the AES encryption and decryption runs in each board
// Child thread can be created by calling creation routine }

Void* child thread()
   {
// again another thread can be created
   }
```

As soon as the Xilkernel library is being called, the scheduler takes the responsibility to execute the thread in a sequential manner depending on type of scheduling imposed, control doesn't return back to main function for specific function call.

As far as the scheduling among the threads is concerned it can be either of the two types that Xilkernel supports, one is Round robin type and another of priority based [18].

### IV. HARDWARE ARCHITECTURAL DESIGN

The proposed work is implemented using the Xilinx EDK 11.1 (version) and Xilinx Spartan 3E FPGA prototyping board has been used for the hardware implementation and testing. Using the Xilinx platform studio from EDK (Embedded Development Kit) the hardware portion of the embedded system has been developed. A soft core 32-bit RISC processor Micro Blaze has been used as a CPU for this embedded computing unit and all the required soft core peripherals are UART 1(used for RS232 DCE (Data Circuit-Terminal Equipment) port), UART 2(used for RS232 DTE (Data Terminal Equipment) port). ). The blocks used to build up the FPGA based embedded computing unit is shown in Fig. 2.

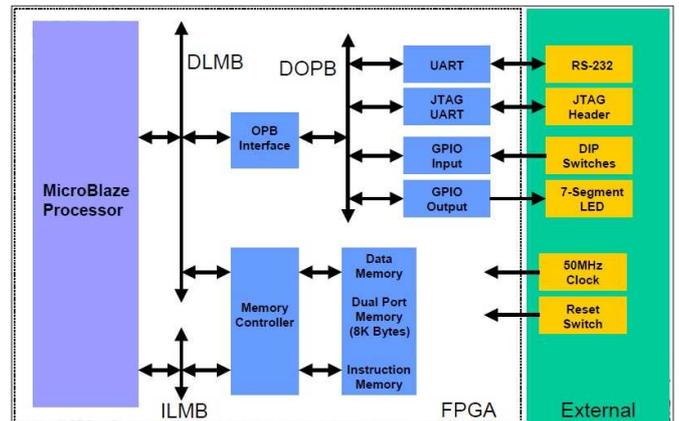
Fig. 2: Internal Architecture Block of FPGA Systems

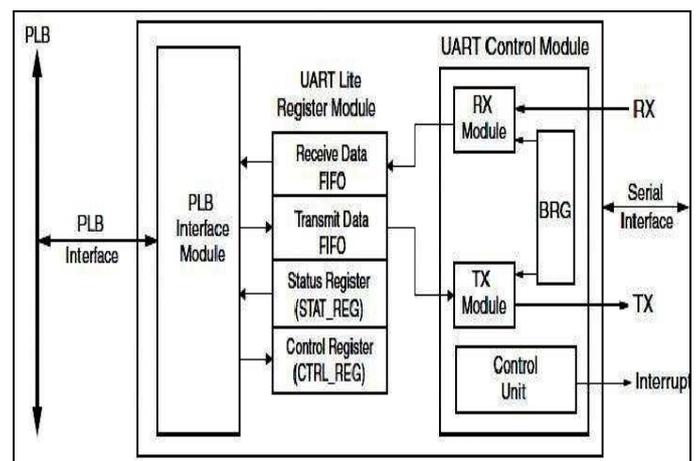
Fig. 3: Universal Asynchronous Receiver Transmitter (UART) system module attached with the Processor Local Bus

## V. SERIAL COMMUNICATION

The component that is used for establishing the serial communication between the multiple FPGA systems is the UART (Universal Asynchronous Receiver Transmitter). The Block diagram of the UART is shown in Fig. 3. Here "BRG" stands for "Baud Rate Generator" which controls the speed of the data communication in RS232 channel. Both receiver and sender side must work in the same band ratio otherwise data will be lost. BRG controls the received data store initially at received FIFO and the *transmit data FIFO* transfer the data through the transmitter Module (TX Module). Status and Control registers are used to check status of the FIFO whether it is full or empty and the control of RX (Receiver) and TX modules respectively.

## VI. IMPLEMENTATION

The proposed architecture was synthesized using Xilinx ISE 11.1 [22] and was implemented on XC3S500e Spartan 3E FPGA Board [22]. The necessary software for this design is written using the feature-rich C/C++ code editor and compilation environment provided within the SDK (Xilinx Software Development Kit). The SDK provides an environment for creating software platforms and applications targeted for Xilinx embedded processor (MicroBlaze). SDK works with hardware designs created with Xilinx Platform Studio (XPS) [22]. We have tested the real time execution of the program in the hardware in RTOS environment. The Fig. 4 gives the work flow of the RTOS.

Xilkernel (RTOS) can be configured by the software platform settings of the EDK. At the time of configuration we can also set the scheduling method, here we chose Round-Robin scheduling with specific time slice. The kernel starts with the execution of a static thread. The real time data are taken from the key board into the board 1 through RS232 port where the AES encryption process is going on. The encrypted plain text called cipher text is being sent to the board 2 using RS232 port. After receiving the cipher text, board 2 decrypts the encrypted text and converts to the plaintext again. This plaintext is being sent to the hyper terminal of a PC also connected with the board 2, through the RS232 interface. Fig. 5 shows the architectural picture of the proposed work.

### A. Results

In our test case we have taken 16 byte data by the key board
*Input = [36 46 e6 a8 88 5a 30 8c 28 31 98 a2 e0 37 07 34].*
*Cipher Key= [2b 7e 15 16 28 ae d2 a6 ab f7 15 88 09 cf 4f 3c].*
Then after ten rounds of the AES the cipher text will appear as
*Cipher text= [27 37 c1 82 83 29 a4 f1 43 93 9c 5e a6 b0 7c e1]* as shown in Fig. 7. For decryption we use cipher text as input and use the same cipher key for decryption algorithm and find *original data= [36 46 e6 a8 88 5a 30 8c 28 31 98 a2 e0 37 07 34].*
This decrypted data is sent to the PC through an RS232 port and been verified using the hyper terminal interface (Fig. 8). Table II shows the resource utilization of the FPGA in our design and Table III shows the execution time of our algorithm and the throughput of encryption and decryption from the equation given below:
*Throughput= (bits processed for encryption or decryption) / second*

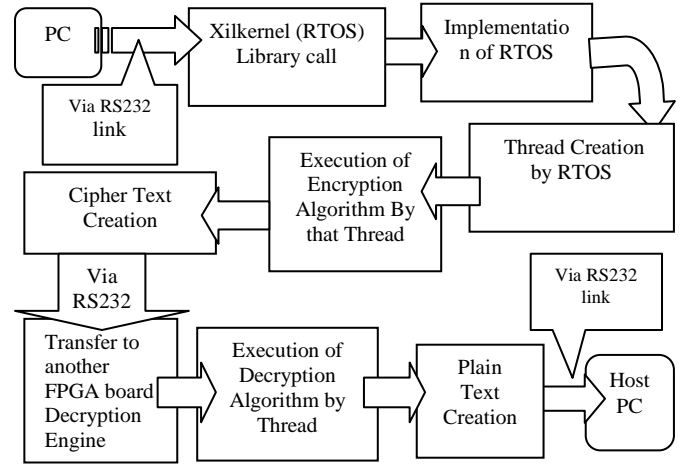

Fig. 4: Work flow using RTOS

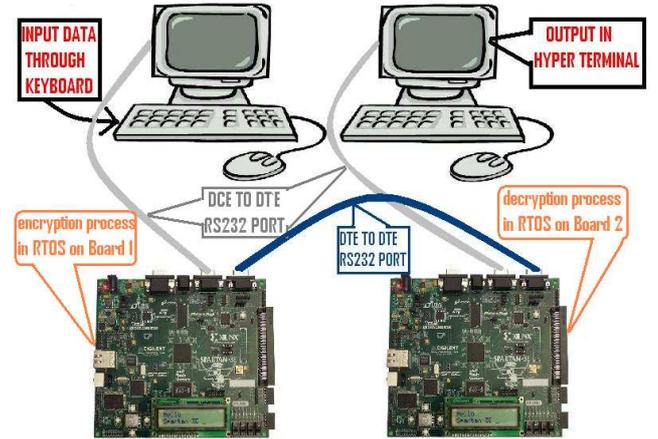

Fig. 5: Architectural Picture of the Experiment.

### B. Comparison with existing works

Table IV, shows the comparison of our work with the existing work and we can claim that our design proposes an efficient solution for the implementation of AES using the FPGA devices.

TABLE II: RESOUCES UNILIZATION

| Logic Utilization | Used | Available | Utilization |
|---|---|---|---|
| Number of Slice Flip Flops | 2,621 | 9,312 | 28% |
| Number of 4 input LUTs | 2,871 | 9,312 | 30% |
| Number of occupied Slices | 2,495 | 4,656 | 53% |
| Number of Slices containing only related logic | 2,495 | 2,495 | 100% |
| Number of Slices containing unrelated logic | 0 | 2,495 | 0% |

TABLE III: ENCRYPTION AND DECRYPTION TIME

| Process | Clock frequency (MHz) | Time (ms) | Throughput (byte/sec) |
|---|---|---|---|
| Encryption | 50 | 4.0274 | 3972.2 |
| Decryption | 50 | 4.1524 | 30825.1 |

TABLE IV: COMPARISION WITH EXSISTING WORKs

| Design | Device | Frequency MHz | Slices | BRAMS |
|---|---|---|---|---|
| Elbirt et al[13] | XCV1000-4 | 31.8 | 10992 | 0 |
| M. McLoone et al[14] | XCV812e-8 | 93.9 | 2000 | 244 |
| K.U.Jarvinen et al[15] | XCV1000e-8 | 129.2 | 11719 | 0 |
| G.P.Saggese [16] | XCV2000e-8 | 158 | 5810 | 0 |
| F. Standaert [17] | XCV3200e-8 | 154 | 15112 | 0 |
| **Proposed** | **Spartan3e Xc3s500e** | **50** | *2,495* | **320** |

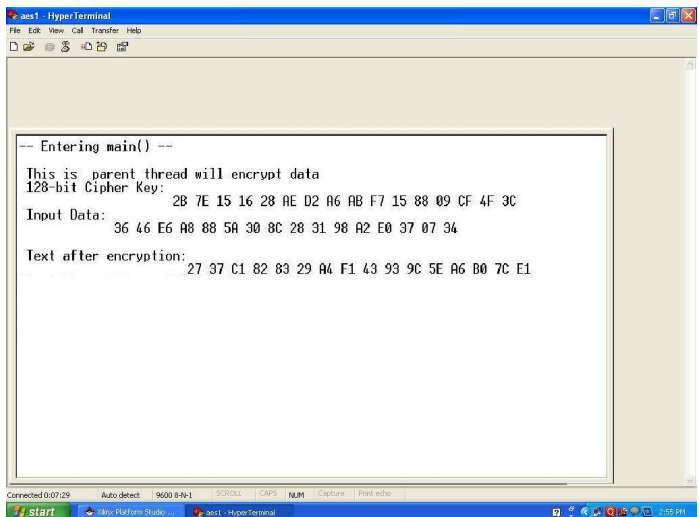
Fig 7: Cipher text on Hyper Terminal

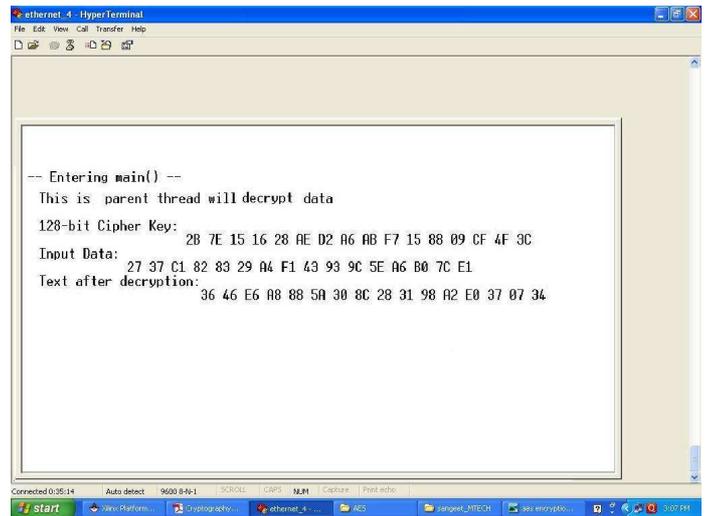
Fig. 8: Decrypted text on Hyper Terminal

VI. CONCLUSION

The aim of this proposed design is to perform a real time data communication exhibiting a significant level of security and providing a faster processing time where necessary. The bit length of the key used in our experiment is 128 bit and the result is obtained successfully. The key width can be varied with a little modification in the algorithm. Though we have performed the real time encryption and decryption of data for RS232 serial communication the technique remains the same for Ethernet data communication using the EMAC core. In future we will try to perform the encryption and decryption of data where inputs will be audio, image, video data coming from different multimedia applications performed over FPGA. Usage of single FPGA with dual processor implementation, where one processor will execute the algorithm while other one will be responsible for input data acquisition, so that the executing processor can handle the algorithm without any interruption, will also be a good step in the world of hardware design.